\documentclass[11pt,twoside]{atmp}
\usepackage{oldgerm}
\newtheorem{Lema}{Lemma}
\newtheorem{Teorema}{Theorem}

\usepackage{amsmath,amssymb}

\usepackage[all]{xy}
\usepackage{color}

\copyrightnotice{2005}{0}{0}{0} %% year, volume, first page, last page.

\begin{document}

\title[A static Einstein metric]
{A static Einstein metric that generalizes the Schwarzschild metric}

\arxurl{0}

\author[Jos\'e L. Mart\'\i nez-Morales]{Jos\'e L. Mart\'\i nez-Morales}

\address{Instituto de Matem\'aticas\\
Universidad Nacional Aut\'onoma de M\'exico\\
A.P. 273, Admon. de correos \#3\\
C.P. 62251 Cuernavaca, Mor.\\
MEXICO}  %lines should be separated with double backslashes: \\
\addressemail{martinez@matcuer.unam.mx}

\begin{abstract}
A static Einstein metric that generalizes the Schwarzschild metric is considered.  The event horizon is not necessarily a sphere and the term $dt\sp2$ is a function on such horizon.  That the metric is Einstein establishes a relation between its terms. One demonstrates that the scalar curvature of the horizon is constant, and that the term $dt\sp2$ gives rise to (i) the metric of the horizon being Einstein, or (ii) the scalar curvature of the horizon being proportional to an eigenvalue of the Laplace operator.
\end{abstract}

\maketitle
\section{Introduction}
We look for metrics that are not Ricci flat, but whose Ricci tensor is proportional to the metric; this condition on a metric defines what is known as an {\it Einstein metric}:
\[
R_{ij}=\Lambda g_{ij}.
\]
The factor $\Lambda$ is necessarily constant, as can be seen from the Bianchi identity for the curvature. In physical terms, when the metric signature is Lorentzian, these are solutions of the vacuum Einstein equations with a cosmological constant $\Lambda$.
\cutpage
As an example, consider the curvature tensor of the Two-Sphere. The Ricci tensor $R_{\mu\nu}$ has the components
\begin{eqnarray*}
R_{\theta\theta}&=&1\\
R_{\theta\phi}&=&0\\
R_{\phi\phi}&=&\sin\sp2\theta.
\end{eqnarray*}
These equations can succinctly be written as
\[
R_{\mu\nu}=g_{\mu\nu},
\]
showing that the standard metric on the two-sphere is an Einstein metric.

Static Einstein metric formulas are used, among others, in:
\begin{itemize}
\item Rotating charged string black holes of space dimension greater than Two \cite{3, 12}.
\item Black holes of String Theory in 4 dimensions \cite{11}.
\item Static electric and magnetic string black holes \cite{11}.
\item The modified Newtonian dynamics paradigm \cite{43}, where the Einstein metric is related to the physical metric by exploiting the direction defined by the gradient of the scalar field (this sidesteps the problem of theories predicting extragalactic gravitational lensing which is too weak if there is indeed no dark matter).
\item Magneto-dilatonic Bianchi-I cosmology, where the properties of the string metric turn out to be quite similar to those of the Einstein metric \cite{Bronnikov}.
\item To compute the thermal stress tensor in a static Einstein space \cite{4}, where the stress tensor is obtained first in a static metric with $g_{00}$ constant, and then the methods of Brown and Cassidy are extended to give it in the conformally related static Einstein metric, where it is conserved and has the correct trace.
\end{itemize}
We consider a static Einstein metric that generalizes the Schwarzschild metric.  The event horizon is not necessarily a sphere and the term $dt\sp2$ is a function on such horizon.  That the metric is Einstein establishes a relation between its terms.  One demonstrates that the scalar curvature of the horizon is constant, and that the term $dt\sp2$ gives rise to (i) the metric of the horizon being Einstein, or (ii) the scalar curvature of the horizon being proportional to an eigenvalue of the Laplace operator. This result does not seem to have been published before.
\section{Statement}
Consider
\begin{itemize}
\item a natural number $n$ greater than Two,
\item two open intervals $I_1$ and $I_2$,
\item an open Euclidean set $S$ of dimension $n$-1,
\item a real and differentiable function $f$ on $I_2$,
\item a real and twice differentiable function $g$ on $I_2$,
\item a real and twice differentiable function $h$ on the Cartesian product of $I_2$ times $S$,
\item a metric {\textgoth h} on $S$, and
\item the (pseudo) metric
\[
\hbox{\textgoth g}=(-) e\sp{2h}dt\otimes dt+e\sp{2f}dx\otimes dx+e\sp{2g}\hbox{\textgoth h}
\]
on the Cartesian product of $I_1$, $I_2$ and $S$.
\end{itemize}
\begin{Teorema}
\label{Teorema}
Suppose that the metric {\textgoth g} is Einstein.  Then,
\begin{enumerate}
\item the scalar curvatures of the metrics {\textgoth g} and {\textgoth h} are constant,
\item there exist
\begin{itemize}
\item a differentiable function $\phi$ on $I_2$, and
\item a differentiable function $\varphi$ on $S$,
\end{itemize}
such that
\[
e\sp h=e\sp{g}(\phi+\varphi),
\]
\item the functions $f$ and $\phi$ can be written in terms of
\begin{itemize}
\item the derivatives of up to first order of the function $g$, and
\item the scalar curvatures of the metrics {\textgoth g} and {\textgoth h},
\end{itemize}
\item either
\begin{enumerate}
\item the function $\varphi$ is an eigenfunction of the Laplace operator in the metric {\textgoth h}, with an eigenvalue that is proportional to the scalar curvature of the metric {\textgoth h}, or
\item
\begin{itemize}
\item the function $h$ is constant on the set $S$, and
\item the metric {\textgoth h} is Einstein.
\end{itemize}
\end{enumerate}
\end{enumerate}
\end{Teorema}
\section{Proof of Theorem \ref{Teorema}}
We write the Ricci tensor of the metric {\textgoth g} in terms of
\begin{itemize}
\item the derivatives of the functions $f$, $g$ and $h$, and
\item the Ricci tensor of the metric {\textgoth h}.
\end{itemize}
We denote by
\begin{eqnarray*}
f'&&\hbox{the derivative function of }f\hbox{ with respect to }x,\\
dh&&\hbox{the differential of }h\hbox{ with respect to the variables of }S,\\
\hbox{Hess}_{\hbox{\textgoth h}}&&\hbox{the Hess operator in the metric {\textgoth h}},\\
\Delta_{\hbox{\textgoth h}}&&\hbox{the Laplace operator in the metric {\textgoth h}},\\
\hbox{Ricci}_{\hbox{\textgoth h}}&&\hbox{the Ricci tensor of the metric {\textgoth h}, and}\\
\hbox{Ricci}_{\hbox{\textgoth g}}&&\hbox{the Ricci tensor of the metric \textgoth g}.
\end{eqnarray*}
\begin{Lema}
\[
\hbox{\rm Ricci}_{\hbox{\textgoth g}}=
\]
\[
(-){e\sp{2 h}}\Big( {e\sp{-2 {g}}} (\Delta_{\hbox{\textgoth h}}h-|dh|\sp2)+ {e\sp{-2 {f}}} \Big(\Big(f\sp{\prime}-(n-1) g\sp{\prime}-{h'}\Big)  {h'}- {h''}\Big)\Big)dt\otimes dt
\]
\[
+\left(-(n-1) {{g\sp{\prime}}\sp2}-(n-1) g''-{{{h'}}\sp2}+ f\sp{\prime} \Big((n-1) g\sp{\prime}+{h'}\Big)- {h''}\right)dx\otimes dx
\]
\begin{equation}
\label{1}
-dx\otimes{e\sp{-h+{g}}}d \Big({e\sp{h-{g}}}\Big)'-{e\sp{-h+{g}}}d \Big({e\sp{h-{g}}}\Big)'\otimes dx
\end{equation}
\[
+\hbox{\rm Ricci}_{\hbox{\textgoth h}}-\hbox{\rm Hess}_{\hbox{\textgoth h}}h-dh\otimes dh+{e\sp{-2 {f}+2 {g}}} \Big( -g''+g\sp{\prime} \Big(f\sp{\prime}-(n-1) g\sp{\prime}-{h'}\Big)\Big)\hbox{\textgoth h}.
\]
\end{Lema}
We now calculate the scalar curvature. We denote by
\begin{eqnarray*}
R_{\hbox{\textgoth g}}&&\hbox{the scalar curvature of the metric {\textgoth g}, and}\\
R_{\hbox{\textgoth h}}&&\hbox{the scalar curvature of the metric {\textgoth h}}.
\end{eqnarray*}
\begin{Lema}
\[
R_{\hbox{\textgoth g}}={e\sp{-2 g}} \Big({R_{\hbox{\textgoth h}}}+2 (\Delta_{\hbox{\textgoth h}}h-|dh|\sp2)\Big)
\]
\begin{equation}
\label{2}
- {e\sp{-2 {f}}} ((n-1){g\sp{\prime}}(n {{g\sp{\prime}}}+ 2 (h'-f'))+ 2 ((n-1) {g''}+{h\sp{\prime}} (h'-f')+{h''})).
\end{equation}
\end{Lema}
\subsection{The term $dt\otimes dt$}
We consider the term $dt\otimes dt$ of the Einstein equation.  Since the scalar curvature of an Einstein metric is constant, we will conclude that
\begin{itemize}
\item the scalar curvature of the metric {\textgoth h} is constant, and that
\item the function $f$ can be written in terms of
\begin{itemize}
\item the derivatives of up to first order of the function $g$, and
\item the scalar curvatures of the metrics {\textgoth g} and {\textgoth h}.
\end{itemize}
\end{itemize}
Suppose that
\begin{equation}
\label{3}
{e\sp{-2 {f}-2 {g}}}\Big( {e\sp{2 {f}}} (\Delta_{\hbox{\textgoth h}}h-|dh|\sp2)+ {e\sp{2 {g}}} \Big(\Big(f\sp{\prime}-(n-1) g\sp{\prime}-{h'}\Big)  {h'}- {h''}\Big)\Big)=\frac{R_{\hbox{\textgoth g}}}{n+1}.
\end{equation}
Multiply (\ref{3}) by 2, and subtract from (\ref{2}) to obtain
\begin{equation}
\label{4}
-(n-1) {R_{\hbox{\textgoth g}}}+{e\sp{-2 g}} (n+1) {R_{\hbox{\textgoth h}}}-{e\sp{-2 {f}}} \big(-1+{n\sp2}\big) \big(n {{{g\sp{\prime}}}\sp2}-2 {g\sp{\prime}} f\sp{\prime}+2 {g''}\big)=0.
\end{equation}
Suppose now that $R_{\hbox{\textgoth g}}$ is constant. Let us solve for the function $f$.
\begin{Lema}$\hbox{}$
\begin{enumerate}
\item$R_{\hbox{\textgoth h}}$ is constant, and
\item a real number $r$ exists so  that
\begin{equation}
\label{5}
e\sp{2f}=-\frac{{e\sp{(n+2) g}} (n-2) (n-1) n (n+1)  {{{g\sp{\prime}}}\sp2}}{{e\sp{2 g}} (n-2) nr+{e\sp{(n+2) g}} (n-2) (n-1) {R_{\hbox{\textgoth g}}}- {e\sp{n g}} n (n+1) {R_{\hbox{\textgoth h}}}}.
\end{equation}
\end{enumerate}
\end{Lema}
\begin{enumerate}
\item It follows from (\ref{4}).
\item Integrate (\ref{4}).
\end{enumerate}
\subsection{The crossed term}
We consider the term (\ref{1}) of the Einstein equation.  Since the metric {\textgoth g} does not have a crossed term, the function $h$ can be written in a particular form.

Suppose that
\[
d \Big({e\sp{h-{g}}}\Big)'=0.
\]
Then, there exist
\begin{itemize}
\item a differentiable function $\phi$ on $I_2$, and 
\item a differentiable function $\varphi$ on $S$,
\end{itemize}
such that
\begin{equation}
\label{6}
e\sp h=e\sp{g}(\phi+\varphi).
\end{equation}
\subsection{The term $dx\otimes dx$}
We consider the term $dx\otimes dx$ of the Einstein equation.  There are two cases:
\begin{enumerate}
\item
\begin{itemize}
\item The function $\phi$ can be written in terms of
\begin{itemize}
\item the function $g$, and
\item the scalar curvatures of the metrics {\textgoth g} and {\textgoth h}, and
\end{itemize}
\item the function $\varphi$ is an eigenfunction of the Laplace operator with eigenvalue the quotient of the scalar curvature $R_{\hbox{\textgoth h}}$ and -2+$n$.
\end{itemize}
\item
\begin{itemize}
\item The function $h$
\begin{itemize}
\item is constant on the set $S$, and
\item can be written in terms of (i) the function $g$, and (ii) the scalar curvatures of the metrics {\textgoth g} and {\textgoth h}, and
\end{itemize}
\item the metric {\textgoth h} is Einstein.
\end{itemize}
\end{enumerate}
Suppose that
\begin{equation}
\label{7}
-(n-1) {{g\sp{\prime}}\sp2}-(n-1) g''-{{{h'}}\sp2}+ f\sp{\prime} \Big((n-1) g\sp{\prime}+{h'}\Big)- {h''}=\frac{R_{\hbox{\textgoth g}}}{n+1}e\sp{2f}.
\end{equation}
\begin{Lema}
\begin{eqnarray}
\label{8}
&&{e\sp{2 g}} (n-2) n {r} \left( (n-2) n{{{g\sp{\prime}}}\sp3}(\phi+\varphi)+ {{\phi}\sp{\prime}} \big((n-4) {{{g\sp{\prime}}}\sp2}+2 {g''}\big)-2 {g\sp{\prime}} {{\phi}''}\right)\nonumber\\
&&+2 {e\sp{n g}} n ( 1+n) {R_{\hbox{\textgoth h}}} \big({{\phi}\sp{\prime}} \big({{{g\sp{\prime}}}\sp2}-{g''}\big)+{g\sp{\prime}} {{\phi}''}\big)\nonumber\\
&&-2 {e\sp{(n+2) g}} ( -n+2) (n-1) {R_{\hbox{\textgoth g}}} \big({{\phi}\sp{\prime}} \big(2 {{{g\sp{\prime}}}\sp2}-{g''}\big)+{g\sp{\prime}} {{\phi}''}\big)=0.
\end{eqnarray}
\end{Lema}
Replace (\ref{5}) and (\ref{6}) in (\ref{7}).
\subsubsection{Case $r$=0}
In this case,
\begin{itemize}
\item the function $\phi$ can be written in terms of
\begin{itemize}
\item the function $g$, and
\item the scalar curvatures of the metrics {\textgoth g} and {\textgoth h}, and
\end{itemize}
\item the function $\varphi$ is an eigenfunction of the Laplace operator with eigenvalue the quotient of the scalar curvature $R_{\hbox{\textgoth h}}$ and -2+$n$.
\end{itemize}

Let us solve for the function $\phi$. By (\ref{8}),
\begin{equation}
\label{9}
n (n+1) {R_{\hbox{\textgoth h}}} \big({{\phi}^{\prime}} \big({{{g^{\prime}}}^2}-{g''}\big)+{g^{\prime}} {{\phi}''}\big)- {e^{2 g}} (n-2) (n-1) {R_{\hbox{\textgoth g}}} \big({{\phi}^{\prime}} \big(2 {{{g^{\prime}}}^2}-{g''}\big)+{g^{\prime}} {{\phi}''}\big)=0.
\end{equation}
Integrate (\ref{9}) to obtain:
\begin{Lema}
A real number $r'$ exists so  that
\begin{equation}
\label{10}
\phi=\frac{{e\sp{-g}} r' {\sqrt{{e\sp{2 g}} (n-2) (n-1) {R_{\hbox{\textgoth g}}}-n (n+1) {R_{\hbox{\textgoth h}}}}}}{n (n+1) {R_{\hbox{\textgoth h}}}}.
\end{equation}
\end{Lema}
\begin{Lema}
The function $\varphi$ is an eigenfunction of the Laplace operator with eigenvalue the quotient of the scalar curvature $R_{\hbox{\textgoth h}}$ and -2+$n$.
\[
(n-2)\Delta_{\hbox{\textgoth h}}\varphi={R_{\hbox{\textgoth h}}}\varphi.
\]
\end{Lema}
Replace (\ref{5}), (\ref{6}) and (\ref{10}) in (\ref{2}).
\subsubsection{Case $r\neq$0}
In this case,
\begin{itemize}
\item the function $h$
\begin{itemize}
\item is constant on the set $S$, and
\item can be written in terms of
\begin{itemize}
\item the function $g$, and
\item the scalar curvatures of the metrics {\textgoth g} and {\textgoth h}, and
\end{itemize}
\end{itemize}
\item the metric {\textgoth h} is Einstein.
\end{itemize}
\begin{Lema}
The function $h$ is constant on the set $S$.
\end{Lema}
By (\ref{8}), $\varphi$ is constant.  By (\ref{6}), $h$ is constant on $S$.

Suppose now that the trace of
\begin{equation}
\label{11}
\hbox{\rm Ricci}_{\hbox{\textgoth h}}-\hbox{\rm Hess}_{\hbox{\textgoth h}}h-dh\otimes dh+{e\sp{-2 {f}+2 {g}}} \Big( -g''+g\sp{\prime} \Big(f\sp{\prime}-(n-1) g\sp{\prime}-{h'}\Big)\Big)\hbox{\textgoth h}-\frac{R_{\hbox{\textgoth g}}}{n+1}e\sp{2g}\hbox{\textgoth h}
\end{equation}
with respect to {\textgoth h} is Zero. Let us solve for the function $h$. Replace (\ref{5}) in the trace of (\ref{11}) with respect to {\textgoth h} to obtain
\begin{equation}
\label{12}
(n-2) n {r} ((n-2) {g\sp{\prime}}+2 {h\sp{\prime}})=2 {e\sp{ng}} \big((n-2) (n-1) {R_{\hbox{\textgoth g}}} ({g\sp{\prime}}-{h\sp{\prime}})+ {e\sp{-2 g}} n (n+1) {R_{\hbox{\textgoth h}}} {h\sp{\prime}}\big).
\end{equation}
Integrate (\ref{12}) to obtain:
\begin{Lema}
A real number $r'$ exists so  that
\begin{equation}
\label{13}
e\sp{2h}=r'\big({e\sp{-(n-2) g}} (n-2) n {r}+{e\sp{2 g}} (n-2) (n-1) {R_{\hbox{\textgoth g}}}-n (n+1) {R_{\hbox{\textgoth h}}}\big).
\end{equation}
\end{Lema}

The metric {\textgoth g} is Einstein if and only if metric {\textgoth h} is Einstein.
\begin{Lema}
\[
\hbox{\rm Ricci}_{\hbox{\textgoth g}}-\frac{R_{\hbox{\textgoth g}}}{n+1}\hbox{\textgoth g}=\hbox{\rm Ricci}_{\hbox{\textgoth h}}-\frac{R_{\hbox{\textgoth h}}}{n-1}\hbox{\textgoth h}.
\]
\end{Lema}
Use (\ref{5}) and (\ref{13}).

\end{document}